\documentclass[fleqn,usenatbib]{mnras}

\usepackage{newtxtext,newtxmath}

\usepackage[T1]{fontenc}

\DeclareRobustCommand{\VAN}[3]{#2}
\let\VANthebibliography\thebibliography
\def\thebibliography{\DeclareRobustCommand{\VAN}[3]{##3}\VANthebibliography}

\usepackage{graphicx}	% Including figure files
\usepackage{amsmath}	% Advanced maths commands
\usepackage{caption,subcaption}     %subfigures

\title[Gaia BH1 and BH2 Progenitor Evolution]
{Gaia BH1 and BH2 -- evolutionary models with overshooting of the black
hole progenitors within the present-day binary separation}

\author[Gilkis \& Mazeh]{
A. Gilkis$^{1}$\thanks{E-mail: agilkis@ast.cam.ac.uk} and T. Mazeh$^{2}$ \\
$^{1}$ Institute of Astronomy, University of Cambridge, Madingley Road, Cambridge CB3 0HA, United Kingdom\\
$^{2}$ School of Physics and Astronomy, Tel Aviv University, Tel Aviv 6997801, Israel\\
}

\date{Accepted 2024 September 23. Received 2024 September 23; in original form 2024 July 12}

\pubyear{\the\year{}}

\begin{document}
\label{firstpage}
\pagerange{\pageref{firstpage}--\pageref{lastpage}}
\maketitle

% Abstract of the paper
\begin{abstract}
Three black holes (BHs) in wide binaries --- Gaia BH1, BH2 and BH3 --- were recently discovered. The likely progenitors of the BHs were massive stars that experienced a supergiant phase, reaching radii of $\sim 1000 R_{\odot}$, before collapsing to form the BH. Such radii are difficult  to accommodate with the present-day orbits of BH1 and BH2 --- with semi-major axes of $1.4$ and $3.7$ au, respectively. In this letter, we show that the maximal radii of the supergiants are not necessarily so large, and realistic stellar evolution models, with some assumed overshooting above the convective core into the radiative stellar envelope, produce substantially smaller maximal radii. The limited expansion of supergiants is consistent with the empirical Humphreys-Davidson limit --- the absence of red supergiants above an upper luminosity limit, notably lower than the highest luminosity of main-sequence stars. We propose that the evolution that led to the formation of Gaia BH1 and BH2 simply did not involve an expansion to the cool supergiant phase.  
\end{abstract}


%%%%%%%%%%%%%%%%% BODY OF PAPER %%%%%%%%%%%%%%%%%%

\section{Introduction}
%=====================

Three dormant black holes (BHs) --- Gaia BH1 \citep{GaiaBH1_23}, BH2 \citep{GaiaBH2_23} and BH3 \citep{GaiaBH3_24}, residing in wide binaries were recently discovered, based on Gaia \citep{Gaia16} astrometric measurements \citep{NSS-DR3-documentation,NSS_23}. The common wisdom assumes that the progenitors of the BHs were probably high-mass stars with masses of $20$--$50$ $M_{\odot}$ that went through a supergiant phase, with radii that could reach up to $\sim 1000 R_{\odot}$ \citep*[e.g.][]{Hurley00}, before collapsing into the singularity of the BH. As discussed thoroughly by \cite{GaiaBH1_23} in their discovery paper of BH1, such radii are difficult to accommodate with the present-day orbits of BH1 and BH2, as their periastron separation is $\sim 150$ and $500 R_{\odot}$, respectively (see details in Table~\ref{tab:orbital_parameters}). One would expect binaries with a separation smaller than the maximal radius of their supergiants to go through a common-envelope phase that results in a dramatic spiral-in of the two components. In such cases, the binary ends up with a short separation and a period on the order of a day \citep[e.g.][]{Antoniadis2022}, similar to most of the known X-ray binaries \citep[e.g.][]{Lutovinov2013,Corral-Santana16}. It is therefore unclear how BH1 and BH2 avoided such a drastic contraction of their binary separation. 

\cite{GaiaBH1_23} and later studies considered a few alternative scenarios for the evolution of BH1 and BH2 \citep[e.g.][]{GaiaBH2_23,MartinPina24}. This includes dynamical capture \citep{rastello23,Di_Carlo24}, interaction with a third distant companion \citep{ perets24} through the eccentricity pumping \citep[][]{MazehShaham79} suggested by \citet{FabryckyTremaine07} as a mechanism for shrinking binary separation, a specific natal kick \citep*{kotko24}, and that the BH itself is a close binary (\citealt{GaiaBH1_23}; \citealt*{hayashi23,BBH_tanikawa24}; but see the recent study of \citealt{nagarajan24}).

An observational hint for another solution to the evolutionary problem of Gaia BH1 and BH2 might come from the empirical absence of cool supergiant stars above a luminosity of $\sim 6\times 10^5\,\mathrm{L}_\odot$, known as the Humphreys-Davidson (hereafter HD) limit (\citealt{HD1979}; see also \citealt*{HDrevisited}). Stars with such a high luminosity have initial masses in the upper-end of the mass range, expected to be progenitors of BHs, and the HD limit is consistent with these high-mass stars not expanding to large radii. \citet{Gilkis2021} considered increased overshooting models in an attempt to explain the HD limit, though other adjustments of mixing parameters might also reproduce this limit. See, for example, \cite{HigginsVink2020} who suggested a combination of {\it highly-efficient} semi-convective mixing \citep[cf.][]{Schootemeijer2019} and {\it decreased} overshooting. While empirical calibrations of the mixing parameters have been done for the typical mass range of massive stars ($10$--$20\,\mathrm{M}_\odot$), there is much more uncertainty for higher masses, which are relevant for BH formation. Regardless of the theoretical explanation of the HD limit, it suggests that the progenitors of BH1 and BH2 did not necessarily reach large radii as assumed.

\cite{GaiaBH1_23} have already raised the possibility that the evolution of Gaia BH1 did not include a giant phase before the BH formation, but noted some difficulties with this scenario (see also \citealt{ElBadry2024} who checked variations in the radial expansion of BH progenitors in the context of Gaia BH3).

In this letter, we show that realistic detailed stellar evolutionary models with some overshooting values produce indeed relatively small maximal radii. Therefore, BH1 and BH2 could have evolved from binaries with an initial separation similar to the present-day orbit or even smaller. Section~\ref{sec:simulation} presents our simulations and Section~\ref{sec:discussion} discusses our results.

%---------------------------------
\begin{table}
	\centering
	\caption{Orbital parameters of the three Gaia BHs.}
	\label{tab:example_table}
	\begin{tabular}{lccccc}
		\hline
		    & $M_2$         & $a$ & $e$ & $a_{\rm peri}$& [Fe/H]   \\
                & [$\mathrm{M}_{\odot}]$ & [au]&  & $[\mathrm{R}_{\odot}]$\\
		\hline
		BH1 & 9.6  & 1.4  & 0.45 & 165  & -0.2\\
		  BH2 & 8.9  & 5.0  & 0.52 & 516  &-0,2\\
		BH3 & 32.7 & 16.2 & 0.73 & 942 &  -2.8\\
		\hline
	\end{tabular}
         \label{tab:orbital_parameters}
\end{table}
%-------------------------------

\section{Simulations}
%=====================
\label{sec:simulation}

In this section we show that the estimate of the maximal expansion of a star strongly depends on the extent of overshooting above the convective core assumed by the model. We simulate rotating single stars with a metallicity of $Z=0.0126$, initial masses between $10\,\mathrm{M}_\odot$ and $135\,\mathrm{M}_\odot$, and an initial equatorial rotation of $200\,\mathrm{km}\,\mathrm{s}^{-1}$, using version 15140 of the Modules for Experiments in Stellar Astrophysics (\textsc{mesa}) code \citep{Paxton2011,Paxton2013,Paxton2015,Paxton2018,Paxton2019}. All models are evolved until the end of the core carbon-burning phase.\footnote{Further details can be found in Appendix~\ref{sec:simdetails} and in a previous study by \cite{Gilkis2021} with similar methodology.} A region of efficient mixing is defined by the Ledoux criterion for convective instability, with radial extension of 
$\alpha H_{\rm P}$ 
above the transition point from instability to stability, where $H_{\rm P}$ is the pressure scale height and $\alpha$ is the overshooting parameter. We take $\alpha=0.335$ from the calibration of \cite{Brott2011} for rotating main-sequence (MS) stars in the mass range of $10$--$20\,\mathrm{M}_\odot$, and follow \cite{Gilkis2021} in considering higher values of $\alpha$, motivated also by indications that overshooting increases with mass (see discussion in Section~\ref{sec:discussion}).

The overshooting extent can have a dramatic effect on the maximal stellar expansion predicted by the evolutionary model, as well as on the final fate of the star  
\citep[e.g.,][]{Eldridge2004,Temaj2024}. This is because the extended mixing region ingests hydrogen-rich fuel into the core, leading to a more massive helium core at the end of the prolonged MS phase. The higher core mass results in a higher stellar luminosity, from the late MS-phase onwards, which increases the radiation-driven wind mass loss. This effect, combined with the longer MS lifetime during which the radiation-driven winds are effective, results in an overall reduction of the leftover hydrogen mass of the post-MS envelope and therefore a smaller radial expansion. 

\begin{figure}
  \centering
    \centering
    \includegraphics[trim=0 0 0 0,clip,width=0.5\textwidth]{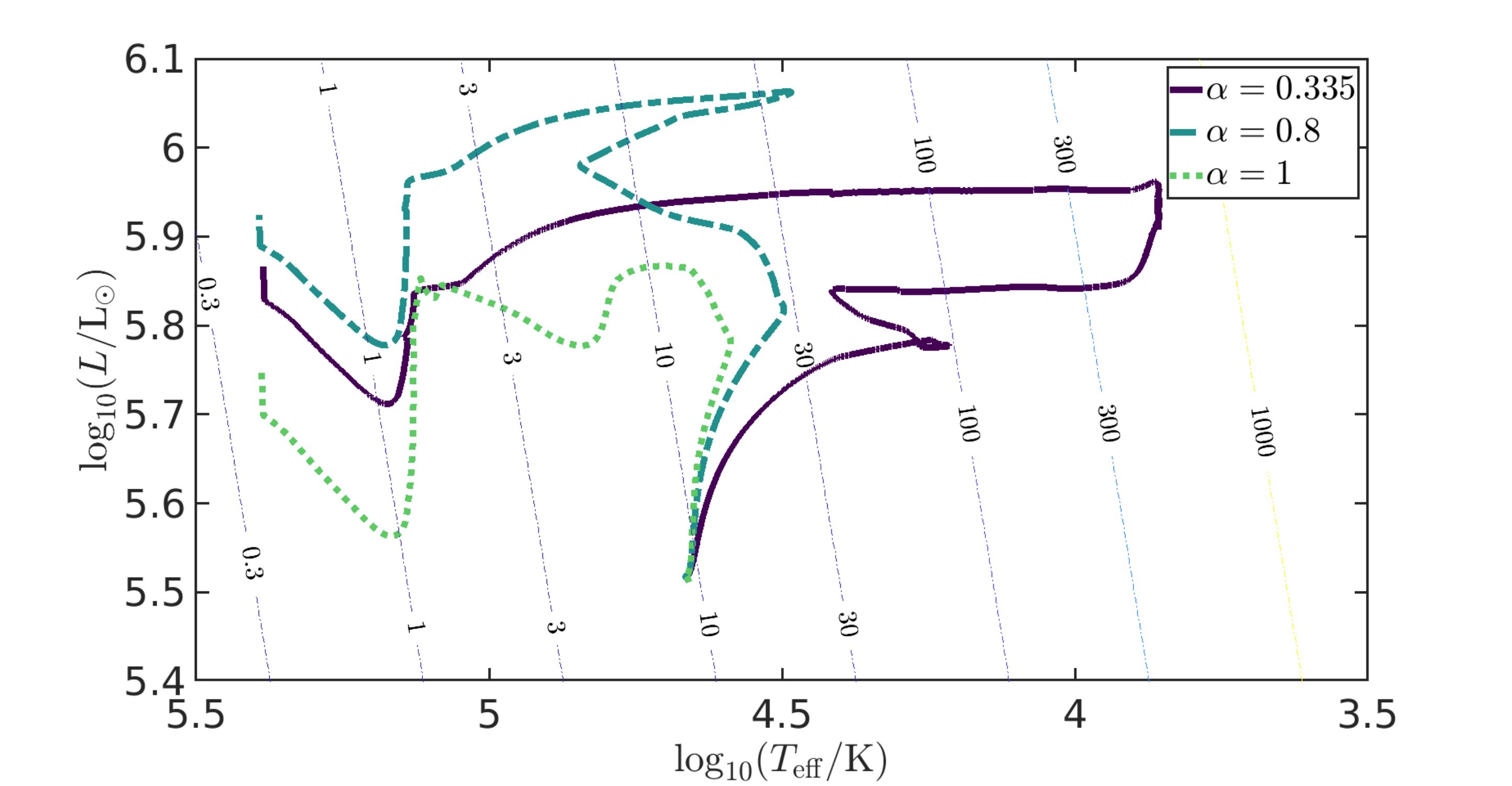}
  \caption{Three stellar evolution tracks with different overshooting parameter $\alpha$ values, for an initial mass of $M_\mathrm{i}= 48\,\mathrm{M}_\odot$ and an initial rotation of $V_\mathrm{i}=200\,\mathrm{km}\,\mathrm{s}^{-1}$, with contours of constant radius.} 
  \label{fig:MESAmodelsHRD}
\end{figure}

In Figure~\ref{fig:MESAmodelsHRD} we show the effect of overshooting on the evolution tracks for a star with an initial mass of $48\,\mathrm{M}_\odot$, 
with three values of the overshooting parameter. One can see that for $\alpha =0.335$ the star expands up to $\ga 500\,\mathrm{R}_\odot$, while for the two higher overshooting models the expansion is limited to $\la 40\,\mathrm{R}_\odot$.

\begin{figure}
    \centering
    \includegraphics[trim=0 0 0 0,clip,width=0.5\textwidth]{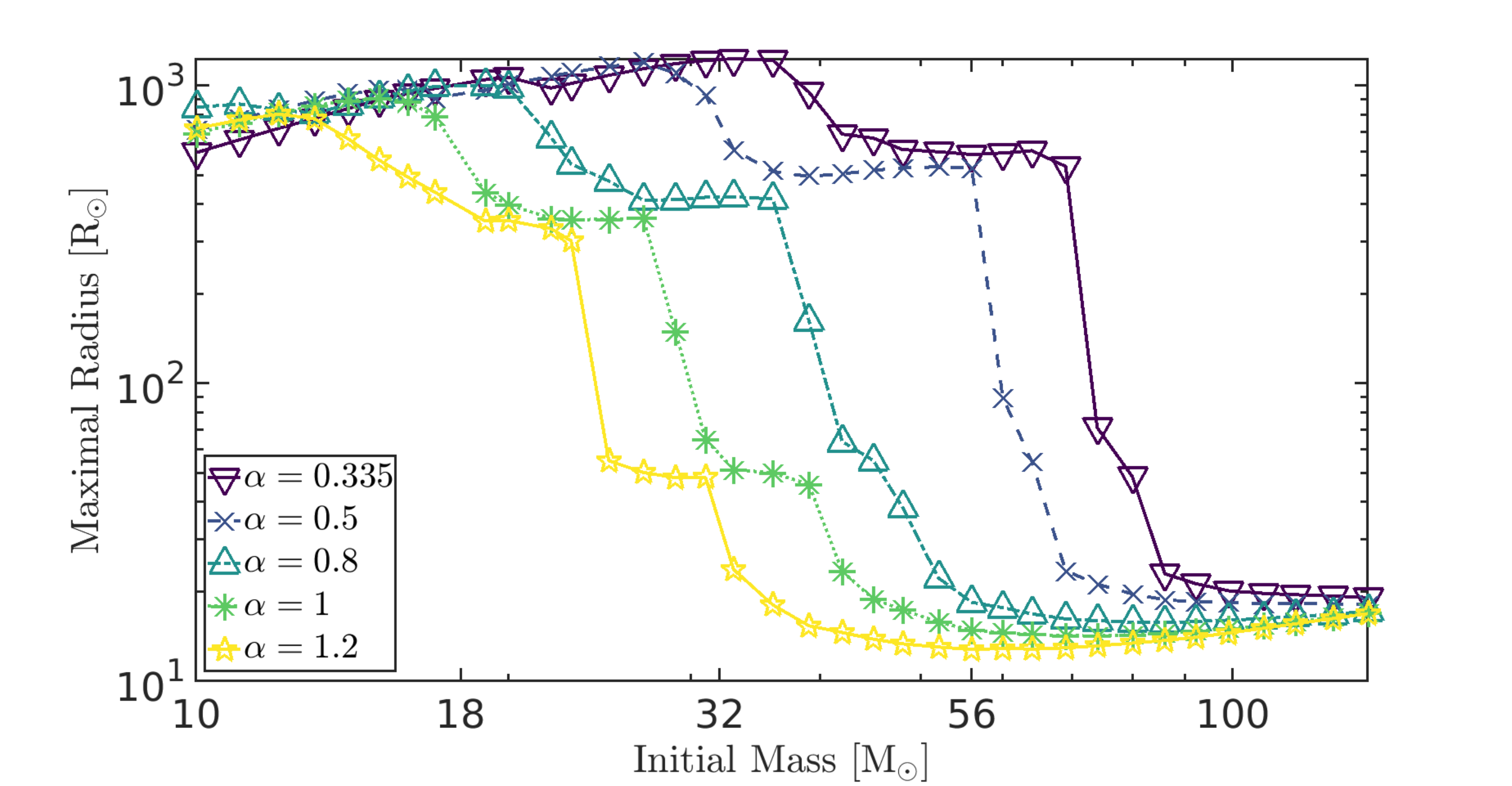}
     \caption{Maximal expansion as a function of the initial mass for evolutionary models with different $\alpha$ values.}
    \label{fig:MESAmodelsMaxR}
%  \end{subfigure}
\end{figure}

In Figure~\ref{fig:MESAmodelsMaxR} we show the effect of overshooting on the {\it maximal} expansion for a few stellar models. The graph of each $\alpha$ value is constructed with $36$ initial masses with equal logarithmic spacing. One can clearly see that the maximal radii can be as small as $\sim 20\, \mathrm{R}_{\odot}$, depending on $\alpha$ and the initial stellar mass.

\begin{figure}
  \centering
    \centering
\includegraphics[trim=0 0 0 0,clip,width=0.5\textwidth]{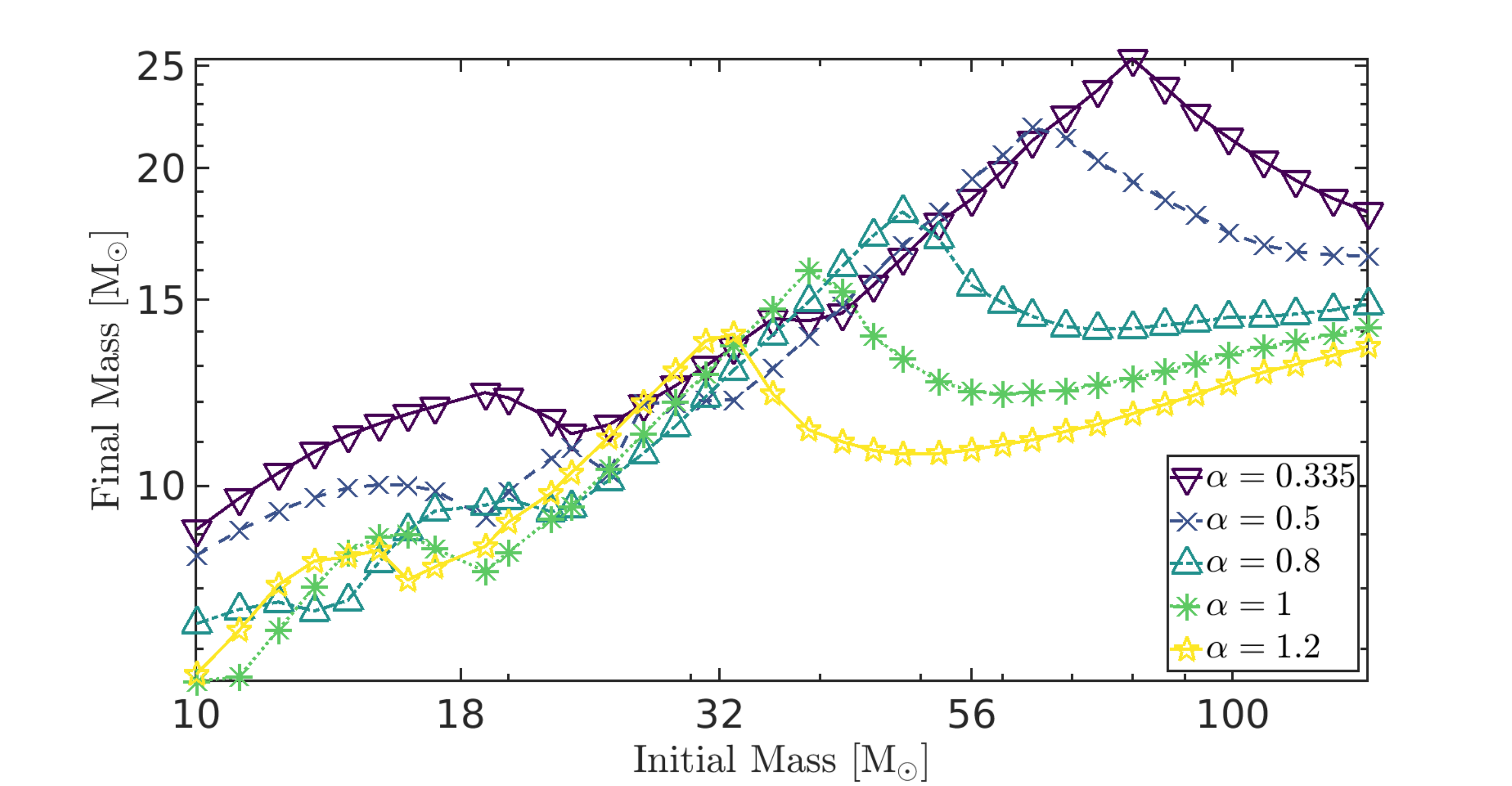}
  \caption{Final mass as a function of the initial mass for evolutionary models with different $\alpha$ values.}
 \label{fig:MESAmodelsFinalMasses}
\end{figure}

In Figure~\ref{fig:MESAmodelsFinalMasses} we present the final stellar mass, defined as the mass at the end of the carbon burning phase, as a function of the initial mass, again for different $\alpha$ values. Note that the final mass is substantially different from the initial one, because of the wind mass loss \citep{VinkNewAR08}. One can see the dramatic impact of the $\alpha$ value on the final mass, for an initial mass of $50\, \mathrm{M}_{\odot}$, for example. One can also notice that the initial mass is not a one-to-one function of the final mass. 
 
\begin{figure}
    \centering
    \centering
    \includegraphics[trim=0 0 0 0,clip,width=0.5\textwidth]{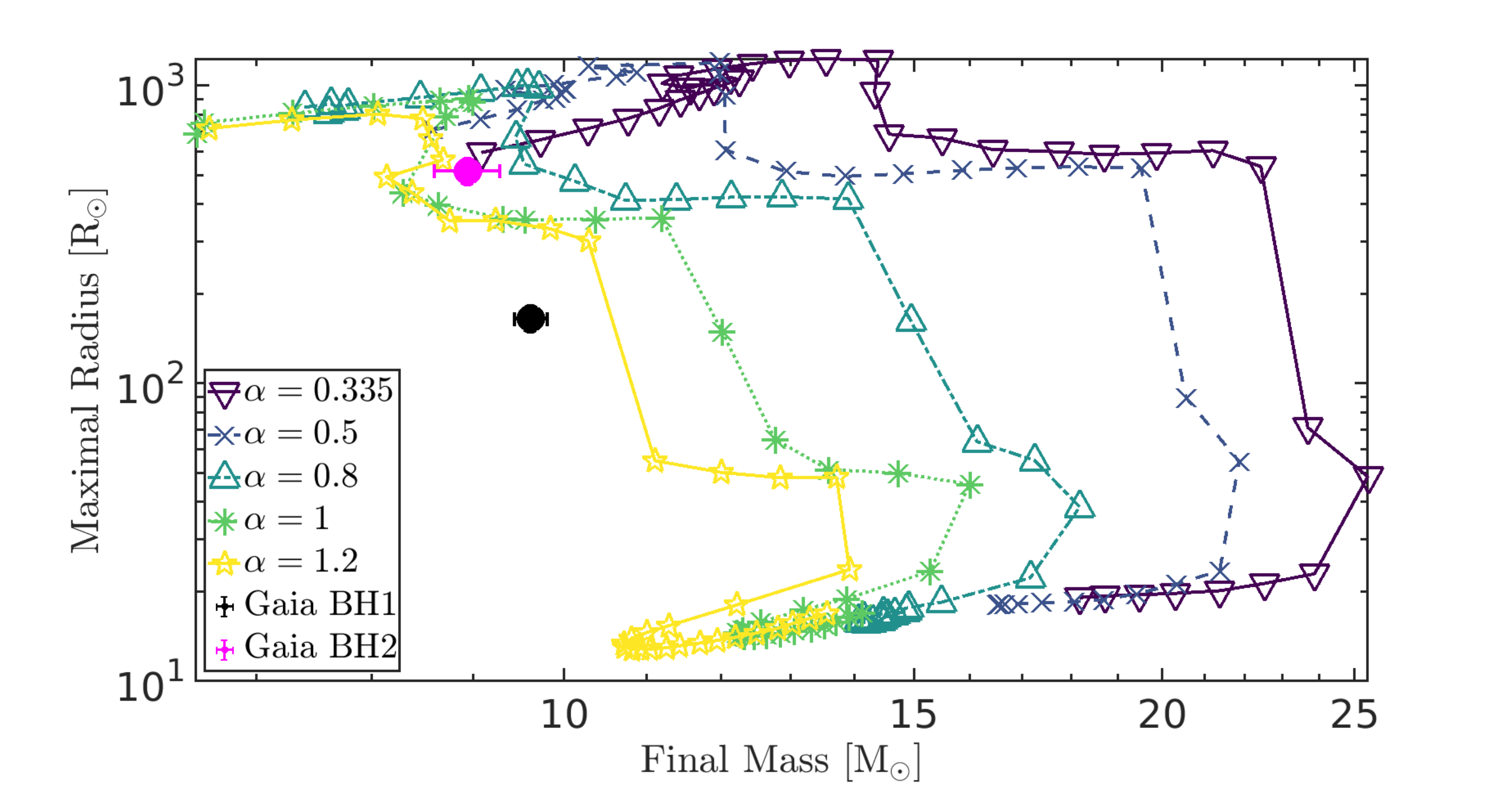}
    \label{fig:MfR300}    
  \caption{Maximal expansion through the stellar evolution models as a function of the final mass. The figure includes the observed masses of BH1 and BH2 at their present-day periastron separation.} 
  \label{fig:MESAmodels}
\end{figure}

Figure~\ref{fig:MESAmodels} shows the maximal radii through the evolutionary track as a function of the final mass as defined for Figure~\ref{fig:MESAmodelsFinalMasses}. We emphasise that this mass can be much larger than the mass of the BH itself, depending on the BH formation processes, which takes place after the stellar evolution phases considered here. The figure includes the observed masses of BH1 and BH2 at their present-day periastron separation. The figure shows that many of the single evolutionary tracks have maximal radii substantially smaller than the observational limits for Gaia BH1 and BH2. Appendix~\ref{sec:simexamples} gives examples of a few detailed radius evolutions.

%---------------------------------------------------
%Figure1
%--------
%------------------------------------------------------

%------------------------------------------------
%Figure2
%---------
%------------------------------------------------------

%----------------------------------------------------
%Figure 3
%---------
%-------------------------------------------------

%------------------------
%Figure 4
%----------
%-------------------------------------------------

\section{Discussion}
%====================
\label{sec:discussion}

We have shown that the maximal radii of massive stars with initial masses of $20$--$50\, \mathrm{M}_{\odot}$ are not necessarily on the order of $1000\, \mathrm{R}_{\odot}$ 
but can be substantially smaller. Models with overshooting layers predict that stars can be confined within the Roche lobe of the present-day orbits of BH1 and BH2 through their whole evolutionary tracks. Therefore, these two systems can be explained by the expansion of the BH progenitors being limited to relatively small radii during their pre-collapse evolution. 

The extent of overshooting in high-mass stars is not well-constrained. Using the observed width of the MS band in the Hertzsprung–Russell diagrams to calibrate the extent of overshooting, \cite{Ekstrom2012} find $\alpha\sim 0.1$, though this calibration was for stars of only a few $\mathrm{M}_\odot$ in mass. The spectroscopic analysis of \cite{Castro2014} shows that the MS widens with increasing luminosity, corresponding to higher masses and therefore indicating a possible increase of overshooting with mass, though statistical limitations in the data prevent detailed quantitative conclusions. \cite{Vink2010} have suggested the possibility of increased overshooting to explain the rotational properties of high-mass stars, and \cite{HigginsVink2019} similarly find that their calibrated rotating massive star models favour $\alpha\sim 0.5$. Observational constraints give a range of values for $\alpha$ up to $\sim 0.5$ \citep{Briquet2007,Johnston2021,AndersPedersen2023}, though these are mostly for stars with masses $\la 16\,\mathrm{M}_\odot$. The calibration of \cite{Brott2011} mentioned above yields $\alpha=0.335$ for a similar mass range. The progenitors of BHs are expected to have higher MS masses, and the extent of overshooting might be higher in those cases. Theoretical studies \citep[e.g.][]{Jermyn2022,Baraffe2023} also suggest that overshooting increases with initial mass. Ongoing studies utilising hydrodynamic simulations of stellar interiors \citep[e.g.][]{Herwig2023,Rizzuti2023,Andrassy2024} and asteroseismology \citep{Aerts2021,Bowman2022,Bowman2023,Shitrit2024} will hopefully enable future stellar evolution simulations to more realistically quantify the extent of overshooting.

Overshooting is not the only uncertain aspect of stellar modelling that affects the expansion of massive stars. \cite{ElBadry2024} experiments with model variations, including overshooting, as well as rotation and the mixing length used within convective regions \citep[see also][]{Chun2018}. Stars with initial masses in the range relevant for BH formation are further plagued by the numerical difficulty in modelling their envelopes when the Eddington luminosity is approached \citep{Agrawal2022a,Agrawal2022b}. Eruptive mass loss caused by super-Eddington regions in massive-star envelopes, which is not usually included in stellar evolution models, might also result in avoiding expansion beyond the HD limit \citep{Cheng2024}.

Neutron star (NS) candidates in wide orbit systems have also been recently discovered \citep{GaiaNS1,NS_El-Badry24}, based on the Gaia astrometry. There is a similar issue in these systems as in Gaia BH1 and BH2, where the expanding progenitor should have overflowed its Roche lobe and initiated common envelope evolution. In the case of NS formation a fast natal kick and significant mass loss during the supernova explosion are expected, and these should be considered in the context of the formation scenarios for these systems, as discussed by \cite{NS_El-Badry24}.

Though we have considered BHs in wide orbits and suggested that limited expansion of the progenitors could explain the inferred lack of interaction, a similar issue exists for shorter period binaries in the context of merging BHs that produce gravitational waves \citep[e.g.][]{Klencki2021}. For example, \cite{Romagnolo2023} explore various stellar modelling approaches leading to different expansions, and conclude that the highest sensitivity is in models with high initial masses that are predicted to expand beyond the HD limit. Therefore, to understand merging BHs, we need first to understand the mechanisms behind this limit.

In conclusion, the new discoveries of dark companions in binaries enabled by Gaia astrometry provide new constraints on evolutionary models of massive stars, with potential implications for understanding the progenitors of core-collapse supernovae, neutron stars and black holes alike.

\section*{Acknowledgements}

We thank an anonymous referee for a constructive review that improved the letter, and Kareem El-Badry for helpful comments. TM thanks the hospitality of Simon Hodgkin and the Kavli Institute for Cosmology visitor programme that enabled a short visit at the Institute of Astronomy and brought about the discussion from which this letter emerged.

%%%%%%%%%%%%%%%%%%%%%%%%%%%%%%%%%%%%%%%%%%%%%%%%%%
\section*{Data Availability}

The input files necessary to reproduce our stellar evolution simulations and associated data products are available at \href{https://zenodo.org/records/13884915}{https://zenodo.org/records/13884915}.

%%%%%%%%%%%%%%%%%%%% REFERENCES %%%%%%%%%%%%%%%%%%

\bibliographystyle{mnras}

\begin{thebibliography}{}
\makeatletter
\relax
\def\mn@urlcharsother{\let\do\@makeother \do\$\do\&\do\#\do\^\do\_\do\%\do\~}
\def\mn@doi{\begingroup\mn@urlcharsother \@ifnextchar [ {\mn@doi@}
  {\mn@doi@[]}}
\def\mn@doi@[#1]#2{\def\@tempa{#1}\ifx\@tempa\@empty \href
  {http://dx.doi.org/#2} {doi:#2}\else \href {http://dx.doi.org/#2} {#1}\fi
  \endgroup}
\def\mn@eprint#1#2{\mn@eprint@#1:#2::\@nil}
\def\mn@eprint@arXiv#1{\href {http://arxiv.org/abs/#1} {{\tt arXiv:#1}}}
\def\mn@eprint@dblp#1{\href {http://dblp.uni-trier.de/rec/bibtex/#1.xml}
  {dblp:#1}}
\def\mn@eprint@#1:#2:#3:#4\@nil{\def\@tempa {#1}\def\@tempb {#2}\def\@tempc
  {#3}\ifx \@tempc \@empty \let \@tempc \@tempb \let \@tempb \@tempa \fi \ifx
  \@tempb \@empty \def\@tempb {arXiv}\fi \@ifundefined
  {mn@eprint@\@tempb}{\@tempb:\@tempc}{\expandafter \expandafter \csname
  mn@eprint@\@tempb\endcsname \expandafter{\@tempc}}}

\bibitem[\protect\citeauthoryear{{Aerts}}{{Aerts}}{2021}]{Aerts2021}
{Aerts} C.,  2021, \mn@doi [Reviews of Modern Physics]
  {10.1103/RevModPhys.93.015001}, \href
  {https://ui.adsabs.harvard.edu/abs/2021RvMP...93a5001A} {93, 015001}

\bibitem[\protect\citeauthoryear{{Agrawal}, {Sz{\'e}csi}, {Stevenson},
  {Eldridge}  \& {Hurley}}{{Agrawal} et~al.}{2022a}]{Agrawal2022a}
{Agrawal} P.,  {Sz{\'e}csi} D.,  {Stevenson} S.,  {Eldridge} J.~J.,   {Hurley}
  J.,  2022a, \mn@doi [\mnras] {10.1093/mnras/stac930}, \href
  {https://ui.adsabs.harvard.edu/abs/2022MNRAS.512.5717A} {512, 5717}

\bibitem[\protect\citeauthoryear{{Agrawal}, {Stevenson}, {Sz{\'e}csi}  \&
  {Hurley}}{{Agrawal} et~al.}{2022b}]{Agrawal2022b}
{Agrawal} P.,  {Stevenson} S.,  {Sz{\'e}csi} D.,   {Hurley} J.,  2022b, \mn@doi
  [\aap] {10.1051/0004-6361/202244044}, \href
  {https://ui.adsabs.harvard.edu/abs/2022A&A...668A..90A} {668, A90}

\bibitem[\protect\citeauthoryear{{Anders} \& {Pedersen}}{{Anders} \&
  {Pedersen}}{2023}]{AndersPedersen2023}
{Anders} E.~H.,  {Pedersen} M.~G.,  2023, \mn@doi [Galaxies]
  {10.3390/galaxies11020056}, \href
  {https://ui.adsabs.harvard.edu/abs/2023Galax..11...56A} {11, 56}

\bibitem[\protect\citeauthoryear{{Andrassy}, {Leidi}, {Higl}, {Edelmann},
  {Schneider}  \& {R{\"o}pke}}{{Andrassy} et~al.}{2024}]{Andrassy2024}
{Andrassy} R.,  {Leidi} G.,  {Higl} J.,  {Edelmann} P.~V.~F.,  {Schneider}
  F.~R.~N.,   {R{\"o}pke} F.~K.,  2024, \mn@doi [\aap]
  {10.1051/0004-6361/202347407}, \href
  {https://ui.adsabs.harvard.edu/abs/2024A&A...683A..97A} {683, A97}

\bibitem[\protect\citeauthoryear{{Antoniadis} et~al.,}{{Antoniadis}
  et~al.}{2022}]{Antoniadis2022}
{Antoniadis} J.,  et~al., 2022, \mn@doi [\aap] {10.1051/0004-6361/202142322},
  \href {https://ui.adsabs.harvard.edu/abs/2022A&A...657L...6A} {657, L6}

\bibitem[\protect\citeauthoryear{{Baraffe} et~al.,}{{Baraffe}
  et~al.}{2023}]{Baraffe2023}
{Baraffe} I.,  et~al., 2023, \mn@doi [\mnras] {10.1093/mnras/stad009}, \href
  {https://ui.adsabs.harvard.edu/abs/2023MNRAS.519.5333B} {519, 5333}

\bibitem[\protect\citeauthoryear{{Bowman}}{{Bowman}}{2023}]{Bowman2023}
{Bowman} D.~M.,  2023, \mn@doi [\apss] {10.1007/s10509-023-04262-7}, \href
  {https://ui.adsabs.harvard.edu/abs/2023Ap&SS.368..107B} {368, 107}

\bibitem[\protect\citeauthoryear{{Bowman} et~al.,}{{Bowman}
  et~al.}{2022}]{Bowman2022}
{Bowman} D.~M.,  et~al., 2022, \mn@doi [\aap] {10.1051/0004-6361/202142375},
  \href {https://ui.adsabs.harvard.edu/abs/2022A&A...658A..96B} {658, A96}

\bibitem[\protect\citeauthoryear{{Briquet}, {Morel}, {Thoul}, {Scuflaire},
  {Miglio}, {Montalb{\'a}n}, {Dupret}  \& {Aerts}}{{Briquet}
  et~al.}{2007}]{Briquet2007}
{Briquet} M.,  {Morel} T.,  {Thoul} A.,  {Scuflaire} R.,  {Miglio} A.,
  {Montalb{\'a}n} J.,  {Dupret} M.~A.,   {Aerts} C.,  2007, \mn@doi [\mnras]
  {10.1111/j.1365-2966.2007.12142.x}, \href
  {https://ui.adsabs.harvard.edu/abs/2007MNRAS.381.1482B} {381, 1482}

\bibitem[\protect\citeauthoryear{{Brott} et~al.,}{{Brott}
  et~al.}{2011}]{Brott2011}
{Brott} I.,  et~al., 2011, \mn@doi [\aap] {10.1051/0004-6361/201016113}, \href
  {https://ui.adsabs.harvard.edu/abs/2011A&A...530A.115B} {530, A115}

\bibitem[\protect\citeauthoryear{{Castro}, {Fossati}, {Langer},
  {Sim{\'o}n-D{\'\i}az}, {Schneider}  \& {Izzard}}{{Castro}
  et~al.}{2014}]{Castro2014}
{Castro} N.,  {Fossati} L.,  {Langer} N.,  {Sim{\'o}n-D{\'\i}az} S.,
  {Schneider} F.~R.~N.,   {Izzard} R.~G.,  2014, \mn@doi [\aap]
  {10.1051/0004-6361/201425028}, \href
  {https://ui.adsabs.harvard.edu/abs/2014A&A...570L..13C} {570, L13}

\bibitem[\protect\citeauthoryear{{Cheng}, {Goldberg}, {Cantiello}, {Bauer},
  {Renzo}  \& {Conroy}}{{Cheng} et~al.}{2024}]{Cheng2024}
{Cheng} S.~J.,  {Goldberg} J.~A.,  {Cantiello} M.,  {Bauer} E.~B.,  {Renzo} M.,
    {Conroy} C.,  2024, \mn@doi [arXiv e-prints] {10.48550/arXiv.2405.12274},
  \href {https://ui.adsabs.harvard.edu/abs/2024arXiv240512274C} {p.
  arXiv:2405.12274}

\bibitem[\protect\citeauthoryear{{Chun}, {Yoon}, {Jung}, {Kim}  \&
  {Kim}}{{Chun} et~al.}{2018}]{Chun2018}
{Chun} S.-H.,  {Yoon} S.-C.,  {Jung} M.-K.,  {Kim} D.~U.,   {Kim} J.,  2018,
  \mn@doi [\apj] {10.3847/1538-4357/aa9a37}, \href
  {https://ui.adsabs.harvard.edu/abs/2018ApJ...853...79C} {853, 79}

\bibitem[\protect\citeauthoryear{{Corral-Santana}, {Casares},
  {Mu{\~n}oz-Darias}, {Bauer}, {Mart{\'\i}nez-Pais}  \&
  {Russell}}{{Corral-Santana} et~al.}{2016}]{Corral-Santana16}
{Corral-Santana} J.~M.,  {Casares} J.,  {Mu{\~n}oz-Darias} T.,  {Bauer} F.~E.,
  {Mart{\'\i}nez-Pais} I.~G.,   {Russell} D.~M.,  2016, \mn@doi [\aap]
  {10.1051/0004-6361/201527130}, \href
  {https://ui.adsabs.harvard.edu/abs/2016A&A...587A..61C} {587, A61}

\bibitem[\protect\citeauthoryear{{Davies}, {Crowther}  \& {Beasor}}{{Davies}
  et~al.}{2018}]{HDrevisited}
{Davies} B.,  {Crowther} P.~A.,   {Beasor} E.~R.,  2018, \mn@doi [\mnras]
  {10.1093/mnras/sty1302}, \href
  {https://ui.adsabs.harvard.edu/abs/2018MNRAS.478.3138D} {478, 3138}

\bibitem[\protect\citeauthoryear{{de Jager}, {Nieuwenhuijzen}  \& {van der
  Hucht}}{{de Jager} et~al.}{1988}]{dJ88}
{de Jager} C.,  {Nieuwenhuijzen} H.,   {van der Hucht} K.~A.,  1988, \aaps,
  \href {https://ui.adsabs.harvard.edu/abs/1988A&AS...72..259D} {72, 259}

\bibitem[\protect\citeauthoryear{{Di Carlo}, {Agrawal}, {Rodriguez}  \&
  {Breivik}}{{Di Carlo} et~al.}{2024}]{Di_Carlo24}
{Di Carlo} U.~N.,  {Agrawal} P.,  {Rodriguez} C.~L.,   {Breivik} K.,  2024,
  \mn@doi [\apj] {10.3847/1538-4357/ad2f2c}, \href
  {https://ui.adsabs.harvard.edu/abs/2024ApJ...965...22D} {965, 22}

\bibitem[\protect\citeauthoryear{{Ekstr{\"o}m} et~al.,}{{Ekstr{\"o}m}
  et~al.}{2012}]{Ekstrom2012}
{Ekstr{\"o}m} S.,  et~al., 2012, \mn@doi [\aap] {10.1051/0004-6361/201117751},
  \href {https://ui.adsabs.harvard.edu/abs/2012A&A...537A.146E} {537, A146}

\bibitem[\protect\citeauthoryear{{El-Badry}}{{El-Badry}}{2024}]{ElBadry2024}
{El-Badry} K.,  2024, \mn@doi [The Open Journal of Astrophysics]
  {10.33232/001c.117652}, \href
  {https://ui.adsabs.harvard.edu/abs/2024OJAp....7E..38E} {7, 38}

\bibitem[\protect\citeauthoryear{{El-Badry} et~al.,}{{El-Badry}
  et~al.}{2023a}]{GaiaBH1_23}
{El-Badry} K.,  et~al., 2023a, \mn@doi [\mnras] {10.1093/mnras/stac3140}, \href
  {https://ui.adsabs.harvard.edu/abs/2023MNRAS.518.1057E} {518, 1057}

\bibitem[\protect\citeauthoryear{{El-Badry} et~al.,}{{El-Badry}
  et~al.}{2023b}]{GaiaBH2_23}
{El-Badry} K.,  et~al., 2023b, \mn@doi [\mnras] {10.1093/mnras/stad799}, \href
  {https://ui.adsabs.harvard.edu/abs/2023MNRAS.521.4323E} {521, 4323}

\bibitem[\protect\citeauthoryear{{El-Badry} et~al.,}{{El-Badry}
  et~al.}{2024a}]{GaiaNS1}
{El-Badry} K.,  et~al., 2024a, \mn@doi [The Open Journal of Astrophysics]
  {10.33232/001c.116675}, \href
  {https://ui.adsabs.harvard.edu/abs/2024OJAp....7E..27E} {7, 27}

\bibitem[\protect\citeauthoryear{{El-Badry} et~al.,}{{El-Badry}
  et~al.}{2024b}]{NS_El-Badry24}
{El-Badry} K.,  et~al., 2024b, \mn@doi [The Open Journal of Astrophysics]
  {10.33232/001c.121261}, \href
  {https://ui.adsabs.harvard.edu/abs/2024OJAp....7E..58E} {7, 58}

\bibitem[\protect\citeauthoryear{{Eldridge} \& {Tout}}{{Eldridge} \&
  {Tout}}{2004}]{Eldridge2004}
{Eldridge} J.~J.,  {Tout} C.~A.,  2004, \mn@doi [\mnras]
  {10.1111/j.1365-2966.2004.08041.x}, \href
  {https://ui.adsabs.harvard.edu/abs/2004MNRAS.353...87E} {353, 87}

\bibitem[\protect\citeauthoryear{{Fabrycky} \& {Tremaine}}{{Fabrycky} \&
  {Tremaine}}{2007}]{FabryckyTremaine07}
{Fabrycky} D.,  {Tremaine} S.,  2007, \mn@doi [\apj] {10.1086/521702}, \href
  {https://ui.adsabs.harvard.edu/abs/2007ApJ...669.1298F} {669, 1298}

\bibitem[\protect\citeauthoryear{{Gaia Collaboration} et~al.,}{{Gaia
  Collaboration} et~al.}{2016}]{Gaia16}
{Gaia Collaboration} et~al., 2016, \mn@doi [\aap]
  {10.1051/0004-6361/201629272}, \href
  {https://ui.adsabs.harvard.edu/abs/2016A&A...595A...1G} {595, A1}

\bibitem[\protect\citeauthoryear{{Gaia Collaboration} et~al.,}{{Gaia
  Collaboration} et~al.}{2023}]{NSS_23}
{Gaia Collaboration} et~al., 2023, \mn@doi [\aap]
  {10.1051/0004-6361/202243782}, \href
  {https://ui.adsabs.harvard.edu/abs/2023A&A...674A..34G} {674, A34}

\bibitem[\protect\citeauthoryear{{Gaia Collaboration} et~al.,}{{Gaia
  Collaboration} et~al.}{2024}]{GaiaBH3_24}
{Gaia Collaboration} et~al., 2024, \mn@doi [\aap]
  {10.1051/0004-6361/202449763}, \href
  {https://ui.adsabs.harvard.edu/abs/2024A&A...686L...2G} {686, L2}

\bibitem[\protect\citeauthoryear{{Generozov} \& {Perets}}{{Generozov} \&
  {Perets}}{2024}]{perets24}
{Generozov} A.,  {Perets} H.~B.,  2024, \mn@doi [\apj]
  {10.3847/1538-4357/ad2356}, \href
  {https://ui.adsabs.harvard.edu/abs/2024ApJ...964...83G} {964, 83}

\bibitem[\protect\citeauthoryear{{Gilkis}, {Shenar}, {Ramachandran}, {Jermyn},
  {Mahy}, {Oskinova}, {Arcavi}  \& {Sana}}{{Gilkis} et~al.}{2021}]{Gilkis2021}
{Gilkis} A.,  {Shenar} T.,  {Ramachandran} V.,  {Jermyn} A.~S.,  {Mahy} L.,
  {Oskinova} L.~M.,  {Arcavi} I.,   {Sana} H.,  2021, \mn@doi [\mnras]
  {10.1093/mnras/stab383}, \href
  {https://ui.adsabs.harvard.edu/abs/2021MNRAS.503.1884G} {503, 1884}

\bibitem[\protect\citeauthoryear{{Hainich} et~al.,}{{Hainich}
  et~al.}{2014}]{Hainich2014}
{Hainich} R.,  et~al., 2014, \mn@doi [\aap] {10.1051/0004-6361/201322696},
  \href {https://ui.adsabs.harvard.edu/abs/2014A&A...565A..27H} {565, A27}

\bibitem[\protect\citeauthoryear{{Hayashi}, {Suto}  \& {Trani}}{{Hayashi}
  et~al.}{2023}]{hayashi23}
{Hayashi} T.,  {Suto} Y.,   {Trani} A.~A.,  2023, \mn@doi [\apj]
  {10.3847/1538-4357/acf4f6}, \href
  {https://ui.adsabs.harvard.edu/abs/2023ApJ...958...26H} {958, 26}

\bibitem[\protect\citeauthoryear{{Herwig} et~al.,}{{Herwig}
  et~al.}{2023}]{Herwig2023}
{Herwig} F.,  et~al., 2023, \mn@doi [\mnras] {10.1093/mnras/stad2157}, \href
  {https://ui.adsabs.harvard.edu/abs/2023MNRAS.525.1601H} {525, 1601}

\bibitem[\protect\citeauthoryear{{Higgins} \& {Vink}}{{Higgins} \&
  {Vink}}{2019}]{HigginsVink2019}
{Higgins} E.~R.,  {Vink} J.~S.,  2019, \mn@doi [\aap]
  {10.1051/0004-6361/201834123}, \href
  {https://ui.adsabs.harvard.edu/abs/2019A&A...622A..50H} {622, A50}

\bibitem[\protect\citeauthoryear{{Higgins} \& {Vink}}{{Higgins} \&
  {Vink}}{2020}]{HigginsVink2020}
{Higgins} E.~R.,  {Vink} J.~S.,  2020, \mn@doi [\aap]
  {10.1051/0004-6361/201937374}, \href
  {https://ui.adsabs.harvard.edu/abs/2020A&A...635A.175H} {635, A175}

\bibitem[\protect\citeauthoryear{{Humphreys} \& {Davidson}}{{Humphreys} \&
  {Davidson}}{1979}]{HD1979}
{Humphreys} R.~M.,  {Davidson} K.,  1979, \mn@doi [\apj] {10.1086/157301},
  \href {https://ui.adsabs.harvard.edu/abs/1979ApJ...232..409H} {232, 409}

\bibitem[\protect\citeauthoryear{{Hurley}, {Pols}  \& {Tout}}{{Hurley}
  et~al.}{2000}]{Hurley00}
{Hurley} J.~R.,  {Pols} O.~R.,   {Tout} C.~A.,  2000, \mn@doi [\mnras]
  {10.1046/j.1365-8711.2000.03426.x}, \href
  {https://ui.adsabs.harvard.edu/abs/2000MNRAS.315..543H} {315, 543}

\bibitem[\protect\citeauthoryear{{Jermyn}, {Anders}, {Lecoanet}  \&
  {Cantiello}}{{Jermyn} et~al.}{2022}]{Jermyn2022}
{Jermyn} A.~S.,  {Anders} E.~H.,  {Lecoanet} D.,   {Cantiello} M.,  2022,
  \mn@doi [\apj] {10.3847/1538-4357/ac5f08}, \href
  {https://ui.adsabs.harvard.edu/abs/2022ApJ...929..182J} {929, 182}

\bibitem[\protect\citeauthoryear{{Johnston}}{{Johnston}}{2021}]{Johnston2021}
{Johnston} C.,  2021, \mn@doi [\aap] {10.1051/0004-6361/202141080}, \href
  {https://ui.adsabs.harvard.edu/abs/2021A&A...655A..29J} {655, A29}

\bibitem[\protect\citeauthoryear{{Klencki}, {Nelemans}, {Istrate}  \&
  {Chruslinska}}{{Klencki} et~al.}{2021}]{Klencki2021}
{Klencki} J.,  {Nelemans} G.,  {Istrate} A.~G.,   {Chruslinska} M.,  2021,
  \mn@doi [\aap] {10.1051/0004-6361/202038707}, \href
  {https://ui.adsabs.harvard.edu/abs/2021A&A...645A..54K} {645, A54}

\bibitem[\protect\citeauthoryear{{Kotko}, {Banerjee}  \& {Belczynski}}{{Kotko}
  et~al.}{2024}]{kotko24}
{Kotko} I.,  {Banerjee} S.,   {Belczynski} K.,  2024, \mn@doi [arXiv e-prints]
  {10.48550/arXiv.2403.13579}, \href
  {https://ui.adsabs.harvard.edu/abs/2024arXiv240313579K} {p. arXiv:2403.13579}

\bibitem[\protect\citeauthoryear{{Lutovinov}, {Revnivtsev}, {Tsygankov}  \&
  {Krivonos}}{{Lutovinov} et~al.}{2013}]{Lutovinov2013}
{Lutovinov} A.~A.,  {Revnivtsev} M.~G.,  {Tsygankov} S.~S.,   {Krivonos} R.~A.,
   2013, \mn@doi [\mnras] {10.1093/mnras/stt168}, \href
  {https://ui.adsabs.harvard.edu/abs/2013MNRAS.431..327L} {431, 327}

\bibitem[\protect\citeauthoryear{{Mar{\'\i}n Pina}, {Rastello}, {Gieles},
  {Kremer}, {Fitzgerald}  \& {Rando Forastier}}{{Mar{\'\i}n Pina}
  et~al.}{2024}]{MartinPina24}
{Mar{\'\i}n Pina} D.,  {Rastello} S.,  {Gieles} M.,  {Kremer} K.,  {Fitzgerald}
  L.,   {Rando Forastier} B.,  2024, \mn@doi [\aap]
  {10.1051/0004-6361/202450460}, \href
  {https://ui.adsabs.harvard.edu/abs/2024A&A...688L...2M} {688, L2}

\bibitem[\protect\citeauthoryear{{Mazeh} \& {Shaham}}{{Mazeh} \&
  {Shaham}}{1979}]{MazehShaham79}
{Mazeh} T.,  {Shaham} J.,  1979, \aap, \href
  {https://ui.adsabs.harvard.edu/abs/1979A&A....77..145M} {77, 145}

\bibitem[\protect\citeauthoryear{{Nagarajan} et~al.,}{{Nagarajan}
  et~al.}{2024}]{nagarajan24}
{Nagarajan} P.,  et~al., 2024, \mn@doi [\pasp] {10.1088/1538-3873/ad1ba7},
  \href {https://ui.adsabs.harvard.edu/abs/2024PASP..136a4202N} {136, 014202}

\bibitem[\protect\citeauthoryear{{Nugis} \& {Lamers}}{{Nugis} \&
  {Lamers}}{2000}]{NL00}
{Nugis} T.,  {Lamers} H.~J.~G.~L.~M.,  2000, \aap, \href
  {https://ui.adsabs.harvard.edu/abs/2000A&A...360..227N} {360, 227}

\bibitem[\protect\citeauthoryear{{Paxton}, {Bildsten}, {Dotter}, {Herwig},
  {Lesaffre}  \& {Timmes}}{{Paxton} et~al.}{2011}]{Paxton2011}
{Paxton} B.,  {Bildsten} L.,  {Dotter} A.,  {Herwig} F.,  {Lesaffre} P.,
  {Timmes} F.,  2011, \mn@doi [\apjs] {10.1088/0067-0049/192/1/3}, \href
  {https://ui.adsabs.harvard.edu/abs/2011ApJS..192....3P} {192, 3}

\bibitem[\protect\citeauthoryear{{Paxton} et~al.,}{{Paxton}
  et~al.}{2013}]{Paxton2013}
{Paxton} B.,  et~al., 2013, \mn@doi [\apjs] {10.1088/0067-0049/208/1/4}, \href
  {https://ui.adsabs.harvard.edu/abs/2013ApJS..208....4P} {208, 4}

\bibitem[\protect\citeauthoryear{{Paxton} et~al.,}{{Paxton}
  et~al.}{2015}]{Paxton2015}
{Paxton} B.,  et~al., 2015, \mn@doi [\apjs] {10.1088/0067-0049/220/1/15}, \href
  {https://ui.adsabs.harvard.edu/abs/2015ApJS..220...15P} {220, 15}

\bibitem[\protect\citeauthoryear{{Paxton} et~al.,}{{Paxton}
  et~al.}{2018}]{Paxton2018}
{Paxton} B.,  et~al., 2018, \mn@doi [\apjs] {10.3847/1538-4365/aaa5a8}, \href
  {https://ui.adsabs.harvard.edu/abs/2018ApJS..234...34P} {234, 34}

\bibitem[\protect\citeauthoryear{{Paxton} et~al.,}{{Paxton}
  et~al.}{2019}]{Paxton2019}
{Paxton} B.,  et~al., 2019, \mn@doi [\apjs] {10.3847/1538-4365/ab2241}, \href
  {https://ui.adsabs.harvard.edu/abs/2019ApJS..243...10P} {243, 10}

\bibitem[\protect\citeauthoryear{{Pourbaix} et~al.,}{{Pourbaix}
  et~al.}{2022}]{NSS-DR3-documentation}
{Pourbaix} D.,  et~al., 2022, {Gaia DR3 documentation Chapter 7: Non-single
  stars}, Gaia DR3 documentation, European Space Agency; Gaia Data Processing
  and Analysis Consortium. Online at
  \href{https://gea.esac.esa.int/archive/documentation/GDR3/index.html}{https://gea.esac.esa.int/archive/documentation/GDR3/index.html},
  id. 7

\bibitem[\protect\citeauthoryear{{Rastello}, {Iorio}, {Mapelli}, {Arca-Sedda},
  {Di Carlo}, {Escobar}, {Shenar}  \& {Torniamenti}}{{Rastello}
  et~al.}{2023}]{rastello23}
{Rastello} S.,  {Iorio} G.,  {Mapelli} M.,  {Arca-Sedda} M.,  {Di Carlo} U.~N.,
   {Escobar} G.~J.,  {Shenar} T.,   {Torniamenti} S.,  2023, \mn@doi [\mnras]
  {10.1093/mnras/stad2757}, \href
  {https://ui.adsabs.harvard.edu/abs/2023MNRAS.526..740R} {526, 740}

\bibitem[\protect\citeauthoryear{{Rizzuti}, {Hirschi}, {Arnett}, {Georgy},
  {Meakin}, {Murphy}, {Rauscher}  \& {Varma}}{{Rizzuti}
  et~al.}{2023}]{Rizzuti2023}
{Rizzuti} F.,  {Hirschi} R.,  {Arnett} W.~D.,  {Georgy} C.,  {Meakin} C.,
  {Murphy} A.~S.,  {Rauscher} T.,   {Varma} V.,  2023, \mn@doi [\mnras]
  {10.1093/mnras/stad1572}, \href
  {https://ui.adsabs.harvard.edu/abs/2023MNRAS.523.2317R} {523, 2317}

\bibitem[\protect\citeauthoryear{{Romagnolo}, {Belczynski}, {Klencki},
  {Agrawal}, {Shenar}  \& {Sz{\'e}csi}}{{Romagnolo}
  et~al.}{2023}]{Romagnolo2023}
{Romagnolo} A.,  {Belczynski} K.,  {Klencki} J.,  {Agrawal} P.,  {Shenar} T.,
  {Sz{\'e}csi} D.,  2023, \mn@doi [\mnras] {10.1093/mnras/stad2366}, \href
  {https://ui.adsabs.harvard.edu/abs/2023MNRAS.525..706R} {525, 706}

\bibitem[\protect\citeauthoryear{{Sabhahit}, {Vink}, {Higgins}  \&
  {Sander}}{{Sabhahit} et~al.}{2022}]{Sabhahit2022}
{Sabhahit} G.~N.,  {Vink} J.~S.,  {Higgins} E.~R.,   {Sander} A. A.~C.,  2022,
  \mn@doi [\mnras] {10.1093/mnras/stac1410}, \href
  {https://ui.adsabs.harvard.edu/abs/2022MNRAS.514.3736S} {514, 3736}

\bibitem[\protect\citeauthoryear{{Schootemeijer}, {Langer}, {Grin}  \&
  {Wang}}{{Schootemeijer} et~al.}{2019}]{Schootemeijer2019}
{Schootemeijer} A.,  {Langer} N.,  {Grin} N.~J.,   {Wang} C.,  2019, \mn@doi
  [\aap] {10.1051/0004-6361/201935046}, \href
  {https://ui.adsabs.harvard.edu/abs/2019A&A...625A.132S} {625, A132}

\bibitem[\protect\citeauthoryear{{Shitrit} \& {Arcavi}}{{Shitrit} \&
  {Arcavi}}{2024}]{Shitrit2024}
{Shitrit} N.,  {Arcavi} I.,  2024, \mn@doi [\aj] {10.3847/1538-3881/ad1514},
  \href {https://ui.adsabs.harvard.edu/abs/2024AJ....167...65S} {167, 65}

\bibitem[\protect\citeauthoryear{{Tanikawa}, {Wang}  \& {Fujii}}{{Tanikawa}
  et~al.}{2024}]{BBH_tanikawa24}
{Tanikawa} A.,  {Wang} L.,   {Fujii} M.~S.,  2024, \mn@doi [arXiv e-prints]
  {10.48550/arXiv.2407.03662}, \href
  {https://ui.adsabs.harvard.edu/abs/2024arXiv240703662T} {p. arXiv:2407.03662}

\bibitem[\protect\citeauthoryear{{Temaj}, {Schneider}, {Laplace}, {Wei}  \&
  {Podsiadlowski}}{{Temaj} et~al.}{2024}]{Temaj2024}
{Temaj} D.,  {Schneider} F.~R.~N.,  {Laplace} E.,  {Wei} D.,   {Podsiadlowski}
  P.,  2024, \mn@doi [\aap] {10.1051/0004-6361/202347434}, \href
  {https://ui.adsabs.harvard.edu/abs/2024A&A...682A.123T} {682, A123}

\bibitem[\protect\citeauthoryear{{Tramper}, {Sana}  \& {de Koter}}{{Tramper}
  et~al.}{2016}]{TSK2016}
{Tramper} F.,  {Sana} H.,   {de Koter} A.,  2016, \mn@doi [\apj]
  {10.3847/1538-4357/833/2/133}, \href
  {https://ui.adsabs.harvard.edu/abs/2016ApJ...833..133T} {833, 133}

\bibitem[\protect\citeauthoryear{{Vink}}{{Vink}}{2008}]{VinkNewAR08}
{Vink} J.~S.,  2008, \mn@doi [\nar] {10.1016/j.newar.2008.06.008}, \href
  {https://ui.adsabs.harvard.edu/abs/2008NewAR..52..419V} {52, 419}

\bibitem[\protect\citeauthoryear{{Vink} \& {Sander}}{{Vink} \&
  {Sander}}{2021}]{VS21}
{Vink} J.~S.,  {Sander} A. A.~C.,  2021, \mn@doi [\mnras]
  {10.1093/mnras/stab902}, \href
  {https://ui.adsabs.harvard.edu/abs/2021MNRAS.504.2051V} {504, 2051}

\bibitem[\protect\citeauthoryear{{Vink}, {de Koter}  \& {Lamers}}{{Vink}
  et~al.}{2000}]{Vink2000}
{Vink} J.~S.,  {de Koter} A.,   {Lamers} H.~J.~G.~L.~M.,  2000, \mn@doi [\aap]
  {10.48550/arXiv.astro-ph/0008183}, \href
  {https://ui.adsabs.harvard.edu/abs/2000A&A...362..295V} {362, 295}

\bibitem[\protect\citeauthoryear{{Vink}, {de Koter}  \& {Lamers}}{{Vink}
  et~al.}{2001}]{Vink2001}
{Vink} J.~S.,  {de Koter} A.,   {Lamers} H.~J.~G.~L.~M.,  2001, \mn@doi [\aap]
  {10.1051/0004-6361:20010127}, \href
  {https://ui.adsabs.harvard.edu/abs/2001A&A...369..574V} {369, 574}

\bibitem[\protect\citeauthoryear{{Vink}, {Brott}, {Gr{\"a}fener}, {Langer}, {de
  Koter}  \& {Lennon}}{{Vink} et~al.}{2010}]{Vink2010}
{Vink} J.~S.,  {Brott} I.,  {Gr{\"a}fener} G.,  {Langer} N.,  {de Koter} A.,
  {Lennon} D.~J.,  2010, \mn@doi [\aap] {10.1051/0004-6361/201014205}, \href
  {https://ui.adsabs.harvard.edu/abs/2010A&A...512L...7V} {512, L7}

\bibitem[\protect\citeauthoryear{{Woosley}}{{Woosley}}{2019}]{Woosley2019}
{Woosley} S.~E.,  2019, \mn@doi [\apj] {10.3847/1538-4357/ab1b41}, \href
  {https://ui.adsabs.harvard.edu/abs/2019ApJ...878...49W} {878, 49}

\bibitem[\protect\citeauthoryear{{Yoon}}{{Yoon}}{2017}]{Yoon2017}
{Yoon} S.-C.,  2017, \mn@doi [\mnras] {10.1093/mnras/stx1496}, \href
  {https://ui.adsabs.harvard.edu/abs/2017MNRAS.470.3970Y} {470, 3970}

\makeatother
\end{thebibliography}
\input{BH1-BH2_overshooting.bbl}

%%%%%%%%%%%%%%%%% APPENDICES %%%%%%%%%%%%%%%%%%%%%

\appendix

\section{Evolutionary Simulation Details}
\label{sec:simdetails}
%===================================

The simulations presented here mostly follow \citet{Gilkis2021}, with an updated wind mass-loss scheme, where the main difference is the inclusion of the wind prescription for very massive stars described by \citet{Sabhahit2022}. The complete mass-loss scheme is as follows. First, the bi-stability jump temperature $T_\mathrm{jump}$ is computed following \cite*{Vink2000,Vink2001}. Below $T_\mathrm{jump}$ the mass-loss rate is the higher between the values of \cite{Vink2001} and \cite*{dJ88}. Above $T_\mathrm{jump}$, the wind mass loss is the higher between between \cite{Vink2001} and \citet{Sabhahit2022} if the surface hydrogen mass fraction is above $0.4$ (with the metallicity dependence given by \citealt{VS21}). For the hot, hydrogen-deficient Wolf-Rayet phase the mass-loss rate follows \cite{NL00} if the hydrogen mass fraction is above $0.1$ and the combination of \citet{Hainich2014} and \citet*{TSK2016} otherwise (as implemented also by \citealt{Yoon2017} and \citealt{Woosley2019}). All hot wind mass-loss schemes are scaled by metallicity according to $\log_{10} (Z/\mathrm{Z}_\odot) = -0.2$ (corresponding to the sub-Solar iron abundance) except the \cite{NL00} prescription in which the metallicity dependence was calibrated to include all elements heavier than hydrogen and helium. 

\section{Examples of Radius Evolution Compared to Periastron Separation}
\label{sec:simexamples}
%===================================

\begin{figure}
  \centering
  \begin{subfigure}{0.5\textwidth}
    \centering
    \includegraphics[trim=0 0 0 0,clip,width=\textwidth]{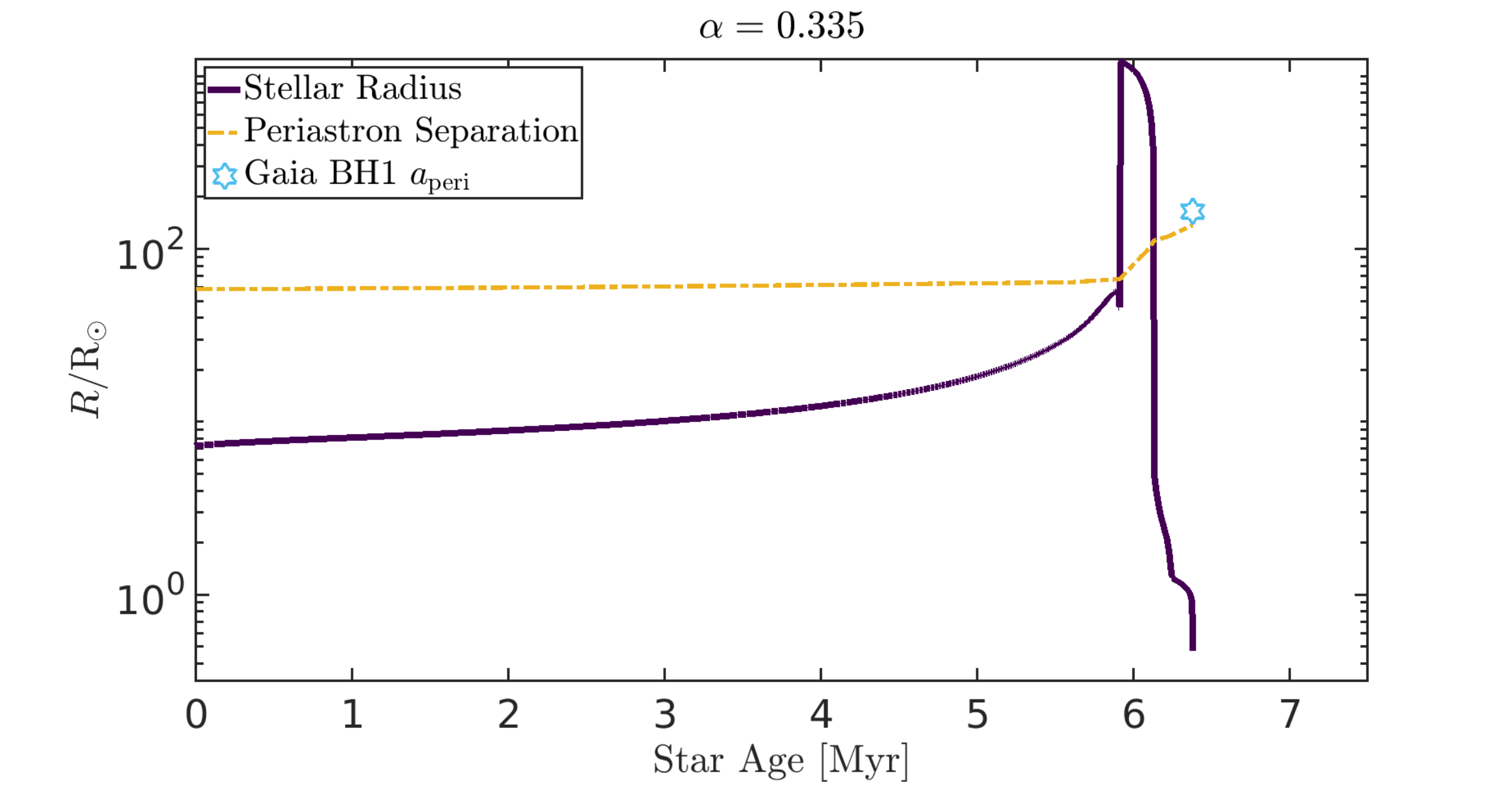}
    \label{fig:M33a0335}
  \end{subfigure}
  \begin{subfigure}{0.5\textwidth}
    \centering
    \includegraphics[trim=0 0 0 0,clip,width=\textwidth]{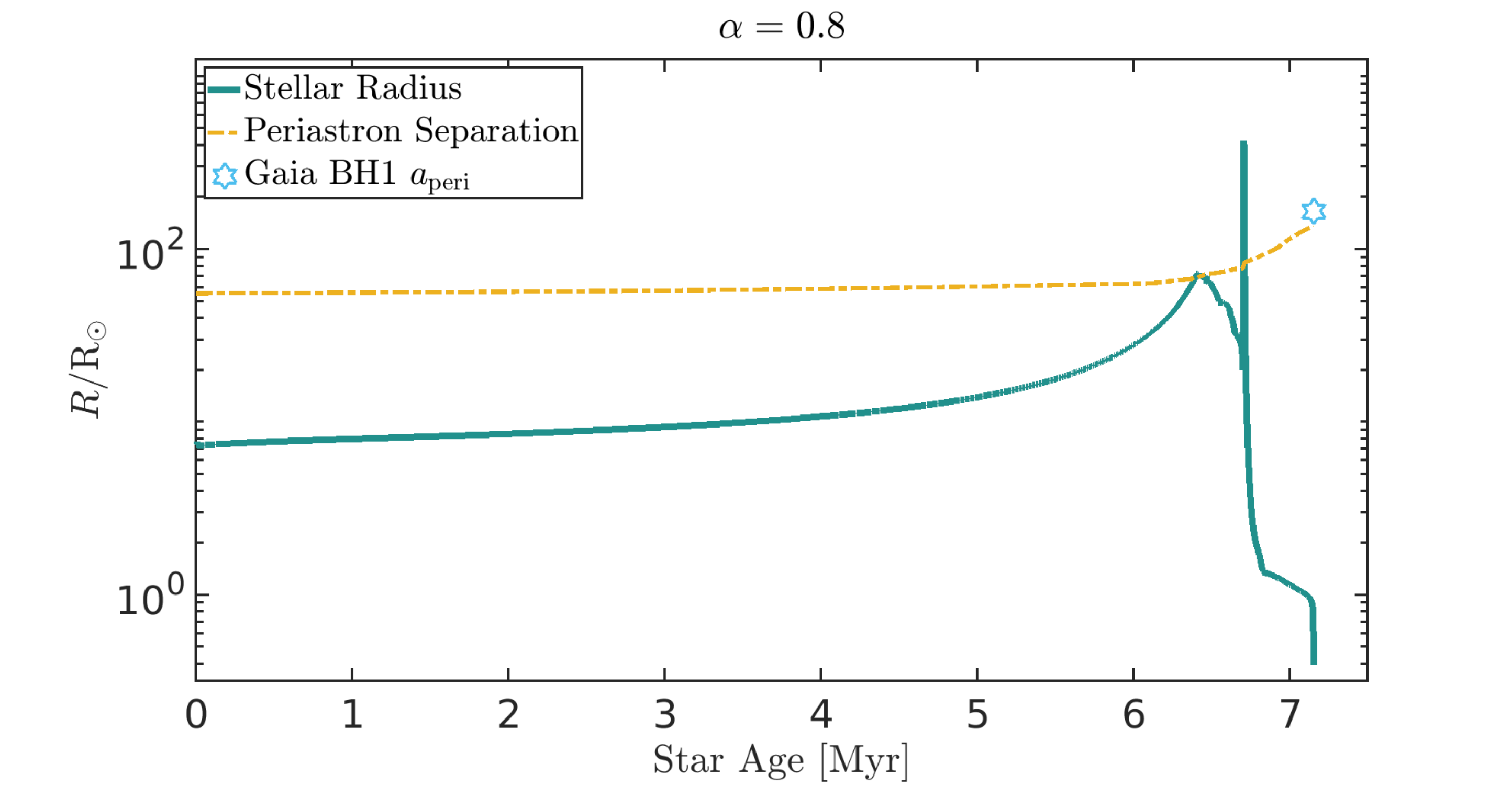}
    \label{fig:M33a0800}
  \end{subfigure}
    \begin{subfigure}{0.5\textwidth}
    \centering
    \includegraphics[trim=0 0 0 0,clip,width=\textwidth]{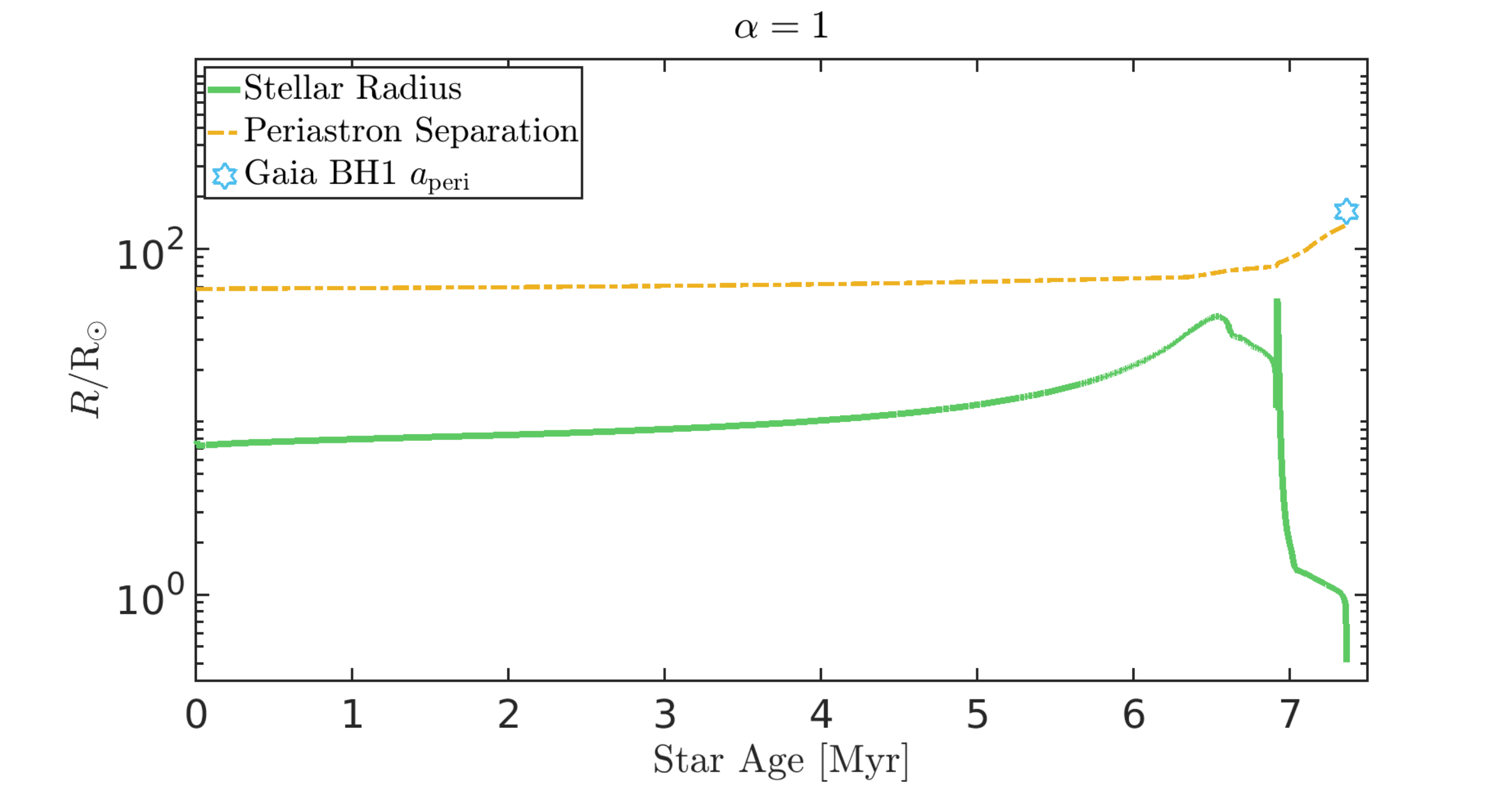}
    \label{fig:M33a1000}
  \end{subfigure}\\    
  \caption{Radius evolution for an initial mass of $33\,\mathrm{M}_\odot$ and an initial rotation of $V_\mathrm{i}=200\,\mathrm{km}\,\mathrm{s}^{-1}$ as a function of time, for overshooting parameter values of $\alpha=0.335$ (top), $\alpha=0.8$ (middle) and $\alpha=1$ (bottom). In each panel, the periastron separation is shown, as computed by reconstructing the orbital parameters prior to the BH formation. The present-day separation of Gaia BH1 is marked at the end of the evolution of the primary star, when it collapses into a BH.} 
  \label{fig:M33R}
\end{figure}

Figure~\ref{fig:M33R} shows three examples of the radius evolution compared to the periastron separation during the evolutionary history of the system prior to the BH formation. The periastron separation is computed by first reconstructing the orbital configuration before the BH formed, assuming an instantaneous mass loss equal to the difference between the mass of the BH in Gaia BH1 and the final mass in the simulations, and a zero-velocity natal kick for the forming BH. This yields the separation and eccentricity just before the BH formation, which were both lower than the present-day values according to our assumptions. For earlier times, the angular momentum of the binary is computed using
\begin{equation*}
J = M_1 M_2 \sqrt{\frac{G a}{M_1 + M_2}}
    \label{eq:B1}
\end{equation*}
and assuming that the angular momentum loss from the stellar wind of the massive primary is isotropic and removes the specific angular momentum of the primary. The wind mass loss of the less massive secondary is negligible, and we assume the eccentricity is constant. These assumptions yield the evolution of the separation as a function of the binary mass,
\begin{equation*}
a \propto (M_1 + M_2)^{-1}.
    \label{eq:B2}
\end{equation*}

Using this reconstruction of the separation in the evolutionary history, we see in Figure~\ref{fig:M33R} that for the present-day separation, the initial separation was smaller, because of the orbit widening as a result of stellar wind mass loss. After the primary star collapses into a BH at the time marked in the plots, the orbit is not affected further by any interaction between the stellar components, and changes to the orbit by the wind mass loss of the low-mass secondary star are negligible. Of the three examples shown in Figure~\ref{fig:M33R}, only the one with $\alpha=1$ complies with the constraints provided by the observations of Gaia BH1.

%%%%%%%%%%%%%%%%%%%%%%%%%%%%%%%%%%%%%%%%%%%%%%%%%%

\bsp	% typesetting comment
\label{lastpage}
\end{document}